\documentclass[referee]{raa}            % referee version: for submission
\usepackage{graphicx,times}             %for PS/EPS graphics inclusion, new
\usepackage{natbib}
\usepackage{amssymb,amsmath}
\bibpunct{(}{)}{;}{a}{}{,}
\usepackage[a4paper=true,dvipdfm=true,pagebackref=true]{hyperref}
\hypersetup{colorlinks = true, linkcolor = green, anchorcolor = red, citecolor = blue, filecolor = red, pagecolor = red, urlcolor = red}

\begin{document}

   \title{Super Penumbral Chromospheric Flare}
   \volnopage{Vol.0 (200x) No.0, 000--000}      %%preserved for Editor. DOn't remove!
   \setcounter{page}{1}          %%starting page, preserved for Editor. DOn't remove!
 \author{S. Liu, H.Q. Zhang
      \inst{1}
   \and D. P. Choudhary
      \inst{1,2}
   \and A. K. Srivastava, B. N. Dwivedi
      \inst{3}
   }


   \institute{National Astronomical Observatories, Chinese Academy of Sciences,
             Beijing 100012, China; {\it debiprasad.choudhary@csun.edu}\\
        \and
             Department of Physics and Astronomy, California State University Northridge\\
18111 Nordhoff St, Northridge, CA, 91330\\
        \and
             Department of Physics
Indian Institute of Technology (BHU), Varanasi-221005, India\\
   }

   \date{Received~~2009 month day; accepted~~2009~~month day}

\abstract{ We observed a C-class flare at the outer boundary of the super-penumbra of a sunspot. The flare was triggered by an emerging magnetic bipolar region that was obliquely oriented with respect to the super-penumbral fibrils. The flare started due to the low height magnetic reconnection of emerging magnetic flux with super-penumbral field resulting hot multi-temperature plasma flows in the inverse Evershed flow channel and its overlying atmosphere. The inverse Evershed flows in the chromosphere start from super penumbra towards sunspot that end at the outer boundary of the penumbra. The hot plasma flow towards the sunspot in the inverse Evershed channels show about 10 km s$^{-1}$ higher velocity in H$\alpha$ wavelengths compared to the plasma emissions at various temperatures as seen in different AIA filters. Even though these velocities are about seven times higher than the typical inverse-Evershed flow speeds, the flow is diminished at the outer boundary of the sunspot's penumbra. This suggests that the super-penumbral field lines that carry the inverse Evershed flows, are discontinued at the boundary where the penumbral field lines dive into the sun and these two sets of field lines are completely distinct. The discontinuity in the typical magnetic field and plasma properties at the adjoining of these two sets of field lines further leads the discontinuity in the characteristic magnetoacoustic and Alfv\'en speeds, therefore, stopping the plasma flows further on. The multi-temperature plasma in the inverse Evershed channels exhibits \textbf{possible} longitudinal oscillations initially during the onset of the flare, and later flows towards the sunspot. In the multi-temperature view, the different layers above the flare region have the mixture of supersonic as well as subsonic flows.
%To the best of our knowledge, we observe firstly that the reconnection generated flare energy release initiates MHD oscillations and multi-temperature plasma flows in super-penumbral regions along with the typical inverse Evershed flows at the chromospheric heights.
}

   \authorrunning{S. Liu et al. }            %author_head in even pages
   \titlerunning{Inverse Evershed Flow from Flare Site}  % title_head in odd pages

   \maketitle
%% The author head (on even pages) and the title head (on odd pages) will be
%% automatically extracted from \author{} and \title{}. Whenever the title is too long,
%% you will be asked to supply a shorter one by inserting either \authorrunning{} or
%% \titlerunning{} before \maketitle. Anyway, you can specify your own heads.
%%
%%
%% Note: In the following text body of your manuscript, please note several differences from
%%       other major journals:
%% (1) \subsection{Please Capitalize the First Letter of Each Notional Word in Subsection Title}
%% (2) Please Capitalize the First Letter of Each Notional Word in all tables' captions

%
%________________________________________________ sections below
%
\section{Introduction}           %% first-level sections will be auto-capitalized
\label{sect:intro}

Solar Flares occur near intense magnetic field regions that contain sunspots. We obtained high resolution chromospheric observations of a solar flare in the superpenumbra of a sunspots that help both understand the trigger mechanism of small flares and structure of sunspot. This paper address both trigger mechanism of small flares and structure of associated sunspot.

Solar flares are most likely caused by a sudden release of magnetic energy due to plasma instabilities or magnetic reconnection (\citeauthor{2005ApJ...629.1135H}, \citeyear{2005ApJ...629.1135H}). For large solar flares it is know that the energy storage can be achieved by twisting the magnetic field due to foot point motion and/or twisted flux emergence and trigger for large flares are due to flux emergence (\citeauthor{1986AdSpR...6....7H}, \citeyear{1986AdSpR...6....7H};
\citeauthor{1998SoPh..179..133C}, \citeyear{1998SoPh..179..133C};
\citeauthor{2004ApJ...611..545G}, \citeyear{2004ApJ...611..545G};
\citeauthor{2008ApJ...687..658W}, \citeyear{2008ApJ...687..658W}).
\citeauthor{2009ApJ...700L.166F} (\citeyear{2009ApJ...700L.166F}) showed that the magnetic field configuration of active regions explodes after achieving their maximum attainable free energy in order to reach the equilibrium state. Several studies have shown the evidence of the chromospheric origin of explosive event activity
(\citeauthor{2005A&A...439.1183}, \citeyear{2005A&A...439.1183};
\citeauthor{2010MmSAI..81..616F}, \citeyear{2010MmSAI..81..616F}).
This is especially true for smaller flares. While large flares are well studied, in case of small flares the trigger mechanisms are their location are not clear. The event presented here gives an opportunity to study the trigger mechanism and its location in chromosphere.

We also use this event to study the sunspot structure by observing the flow of hot flare plasma along the fine structures around the sunspots. Sunspots consist of magnetic flux bundles that protrude through the photosphere and expand into the chromosphere and corona. The magnetic field lines in the umbra
are mostly vertical and inclined to various degree in penumbra and super-penumbra floating above the photosphere in uncombed fashion
(\citeauthor{2000AandA?361..734P}, \citeyear{2000AandA?361..734P}).  %(Martinez Pillet, 2000).
The penumbral-fibrils carry Evershed flows that mostly terminate in a ring of down-flow channels at the outer edge of the boundary with quiet sun (\citeauthor{1999aanda?608.1148P}, \citeyear{1999aanda?608.1148P}).   %(Schlichenmaier and Shimdt, 1999).
About a 10th of penumbral field lines extend beyond the boundary carrying the Evershed flow material and rise up into the super-penumbral magnetic canopy that surrounds a sunspot some 300 km above the photosphere%(Westendorp et al., 1997, Solanki, Ruedi, Livingston, 1992, Ruedi, Solanki, Rabin, 1992).
(\citeauthor{1997Nature?389..47P}, \citeyear{1997Nature?389..47P};
\citeauthor{1992A&A...263..312S}, \citeyear{1992A&A...263..312S};
\citeauthor{1992apj?261.21P}, \citeyear{1992apj?261.21P}).
 Extensive studies of photospheric Evershed Flow along the penumbral fibrils show that the inclined magnetic field from the spot mostly dive down into the photosphere at penumbral boundary, but its connection to the superpenumbra is less clear primarily due to the paucity of chromospheric inverse Evershed flow. These flows are carried by the fibrils that exist in superpenumbra. The photospheric region corresponding to the super-penumbra are observed with moat flows and moving magnetic features MMFs along the direction of penumbral filaments. The super-penumbral fibrils are observed in the chromosphere, carrying inverse Evershed flow material towards the spot at speeds of about 8-10 km s$^{-1}$ (\citeauthor{2003anda?403.1123P}, \citeyear{2003anda?403.1123P}). %(Georgakilas et al, 2003).
Most super-penumbral fibrils terminate at the boundary of penumbra-quiet sun and about one third of them begin inside the umbra
(\citeauthor{2014anda?570.92P}, \citeyear{2014anda?570.92P};
\citeauthor{2004apj?349L.37P}, \citeyear{2004apj?349L.37P}). %(Louis et al, 2014, Balasubramaniam, Pevtsov, Rogers, 2004).

In the three dimensional solar atmosphere, the super-penumbral filament may form the upper boundary layers of the associated canopy, therefore, it may lie in the upper chromosphere and transition region. There are only few reports in the past which describe the inverse Evershed flows at the transition region temperature
 (\citeauthor{2008A&A...491L...5T}, \citeyear{2008A&A...491L...5T}),
while most of the flows are observed in the lower solar chromosphere
(\citeauthor{2012ApJ...750...22V}, \citeyear{2012ApJ...750...22V}).
Recently, an evidence for an accelerated flow ($\approx$40 m s$^{-2}$) along an fibrils anchored at its endpoints in the outer boundary of the sunspot and weaker plage supports the magnetic siphon flows and its role in the formation of the inverse Evershed effect in solar chromosphere (\citeauthor{2013ApJ...768..111S}, \citeyear{2013ApJ...768..111S}).
\citeauthor{2014ApJ...788..183B} (\citeyear{2014ApJ...788..183B})
have found an evidence that the inverse Evershed flows occur into the sunspot in the lower chromosphere due to the siphon flows along short quiescent loops.

These observations suggest that most of the sunspot inclined field channels dive into the photosphere carrying down-flow plasma at the boundary of penumbra and super-penumbra. The structure of magnetic field arrangement in this region, therefore, an important aspect of overall sunspot structure as penumbral and super-penumbral structure could contain separate sets of field bundles that perform mass transfer due to different mechanisms. The magnetic field structuring have been studied in details in the outer periphery of the spots, which may provide more detailed information about the flows there. There are various views on it, {\it viz.}, \citeauthor{2015SoPh..290.1607S} (\citeyear{2015SoPh..290.1607S}) have shown that the superpenumbral magnetic field does not appear to be finely structured, unlike the observed intensity structures indicating that the fibrils are not the concentrations of magnetic fluxes instead they are distinguished by localized thermalization. On the other hand, the actual sunspot penumbra is considered to be inter-combed and fine structured magnetic fields due to the downward pumping of the fluxes
(\citeauthor{2002Natur.420..134T}, \citeyear{2002Natur.420..134T}).
 Normal Evershed effect is associated with the magneto convection and the sunspot penumbral regions
(\citeauthor{2007PASJ...59S.593I}, \citeyear{2007PASJ...59S.593I}).
It was found that the weaker dark regions in-between penumbral filaments of the spot are the likely regions for the upward Evershed flows in the lower solar chromosphere
(\citeauthor{2007ApJ...668L..91B}, \citeyear{2007ApJ...668L..91B}).
Therefore, in the spot's penumbra, this effect may be a nonlinear magnetoconvection that has the properties of traveling waves in the presence of a strong, highly inclined magnetic fields
(\citeauthor{2009ApJ...700L.178K}, \citeyear{2009ApJ...700L.178K}).

Keeping in views that the super penumbra is not a  mere fine structured magnetic tubes instead it is the thermalized atmosphere supplemented by a bulk magnetic field which consists of the outward (e.g., localized jets) and inverse plasma flows (e.g., inverse Evershed effects). Study of such plasma dynamics and their drivers become significant in such regions in hydrodynamic (or magnetohydrodynamic) point-of-views, which circulate mass and energy. As stated above, there are only few studies, especially in the solar chromosphere, which have taken into account the physical processes involved in the triggering of the inverse Evershed flows in the superpenumbral regions.

To the best of our knowledge, this is the first attempt to unveil the multi-temperature view of the inverse Evershed flows in the superpenumbral regions around a sunspot to understand the underlying physical processes in their formation. In this paper, we probe this region by studying the flow of heated plasma resulting from a c-class flare at the outer super-penumbral boundary. Section 2 describes the observational data and its analyses. We outline the observational results and their interpretation in Section 3. Last section depicts the discussion and conclusions of this paper.

%%%%%%%%%%%%%%%%%%%%%%%%%%%%%%%%%%%%%%%%%%%%%%%%%%%%%%%%%%%%%%%%%%%%%%%%%%%%%%%%%%%%%%%%%%%%%%%%%%%%%%%%%%%%%%%%%%%%%%%%%%%%%%%%%%%%%%%%%%%%%%%%
\begin{figure}
%%%   jzhc_check_bzbt_together
   \centerline{\includegraphics[width=1.2\textwidth,clip=]{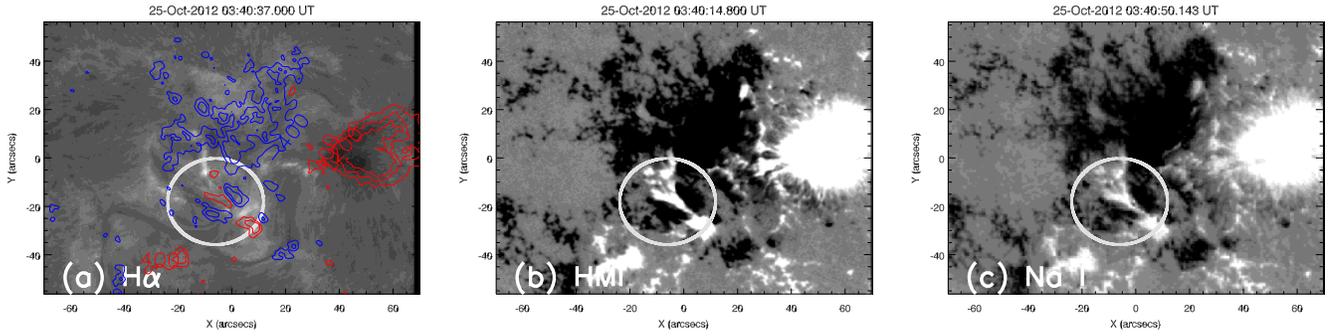}}
  % \caption{The magnetogram and 304\AA~image with sub-regions labeled 1, 2 and 3 to show the magnetic flux and radiation intensity in individual sub-regions.} \label{Fig4}
    \caption{Left-panel (a): H$\alpha$ imageobtained by a New Vacuum Solar Telescope (NVST) with LOS magnetic field contours on 25 October 2012 at 03:40 UT.
    Middle panel (b): HMI LOS magnetic field where red is positive while blue is the negative. Right panle (c): Na I line observation indicating LOS magnetic field similar to that of HMI. White circle drawn in (b) and (c) is the region-of-interest where the inverse Evershed flows are identified.}\label{Fig1}
\end{figure}

%%%%%%%%%%%%%%%%%%%%%%%%%%%%%%%%%%%%%%%%%%%%%%%%%%%%%%%%%%%%%%%%%%%%%%%%%%%%%%%%%%%%%%%%%%%%%%%%%%%%%%%%%%%%%%%%%%%%%%%%%%%%%%%%%%%%%%%%%%%%%%%%%

%%%%%%%%%%%%%%%%%%%%%%%%%%%%%%%%%%%%%%%%%%%%%%%%%%%%%%%%%%%%%%%%%%%%%%%%%%%%%%%%%%%%%%%%%%%%%%%%%%%%%%%%%%%%%%%%%%%%%%%%%%%%%%%%%%%%%%%%%%%%%%%%
\begin{figure}
%%%   jzhc_check_bzbt_together
   \centerline{\includegraphics[width=1.1\textwidth,clip=]{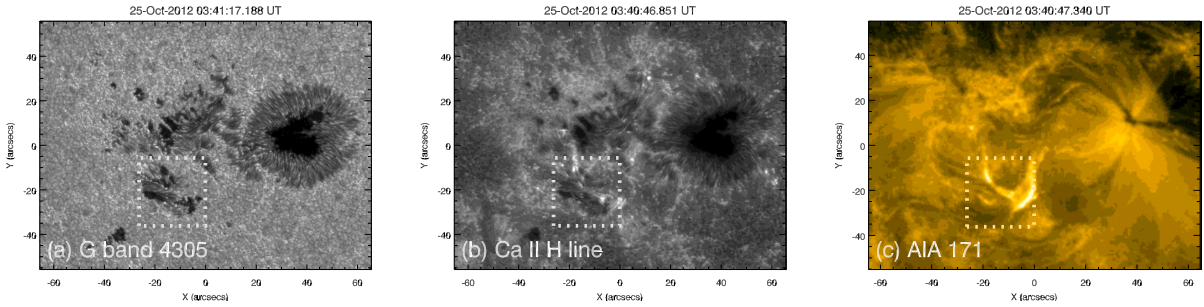}}
  % \caption{The magnetogram and 304\AA~image with sub-regions labeled 1, 2 and 3 to show the magnetic flux and radiation intensity in individual sub-regions.} \label{Fig4}
    \caption{G-band image (left-panel) and Ca II H image (middle-panel) obtained with the Solar Optical Telescope onboard Hinode mission, and AIA 171 \AA image (right-panel) observed by SDO. Brightness is observed at the polarity inversion channel of emerging flux region. The Ca II image shows dark pore structure at the same location.}\label{Fig2}
\end{figure}

%%%%%%%%%%%%%%%%%%%%%%%%%%%%%%%%%%%%%%%%%%%%%%%%%%%%%%%%%%%%%%%%%%%%%%%%%%%%%%%%%%%%%%%%%%%%%%%%%%%%%%%%%%%%%%%%%%%%%%%%%%%%%%%%%%%%%%%%%%%%%%%%%

%% Authors can give a citation as 'Michel et al. 1992'.
%% You may also use \cite, \citep and \citet for citation, and use Table~1 or Figure~1
%% and so forth. Using \ref and \label for cross-references of Tables/Figures
%% is a good way in adjusting/adding/removing text, tables or figures.

\section{Observations and Analysis}
\label{sect:Obs}

The event studied in this paper is a c-class flare occurred at 3:38 UT in NOAA 11598 located at S12E15 on 2012 October 25. In this study, we utilize time-lapse high resolution images through H$\alpha$ filter, and images in ultraviolet wavelengths formed at transition region heights above the sunspot. The H$\alpha$ observations are obtained with a New Vacuum Solar Telescope (NVST) located at Fuxian lake Solar Observatory (FSO) in southwest China. The telescope with its 985 mm aperture was designed by a modified Gregorian system, of which the effective focal length is 45 m  (\citeauthor{2014RAA....14..705L}, \citeyear{2014RAA....14..705L}). The main goal of NVST is to reveal fine structures in both the photosphere and chromosphere through spectral and imaging observations with high spectral and spatial resolution in the wavelength range from 0.3 to 2.5 micron. The NVST has mainly provided H$\alpha$ (656.3 with passband 0.025 nm) and TiO (705.8 with passband 0.01 nm) data for observing
the chromosphere and photosphere, respectively. Generally, the observational data are processed from Level 0 to Level 1 through dark current subtraction and flat field correction, then reconstructed to Level 1 + by the method of speckle masking
(\citeauthor{1977OptCo..21...55W}, \citeyear{1977OptCo..21...55W}). After image reconstruction, H$\alpha$ image has the resolution of 0.162 arcsec/pixel and cadence of 12s for scientific analysis. The processed sequence of H$\alpha$ images are used to make a movie that can be seen in event.mpg. The high resolution image of the first frame show the umbra, penumbra and super-penumbra structure around the sunspot. The movie shows that following a flare at the boundary of super-penumbra, the H$\alpha$ emission moves towards the spot along the super-penumbral fibrils and stop at the penumbral edge.

%%%%%%%%%%%%%%%%%%%%%%%%%%%%%%%%%%%%%%%%%%%%%%%%%%%%%%%%%%%%%%%%%%%%%%%%%%%%%%%%%%%%%%%%%%%%%%%%%%%%%%%%%%%%%%%%%%%%%%%%%%%%%%%%%%%%%%%%%%%%%%%%
\begin{figure}
%%%   jzhc_check_bzbt_together
   \centerline{\includegraphics[width=1\textwidth,clip=]{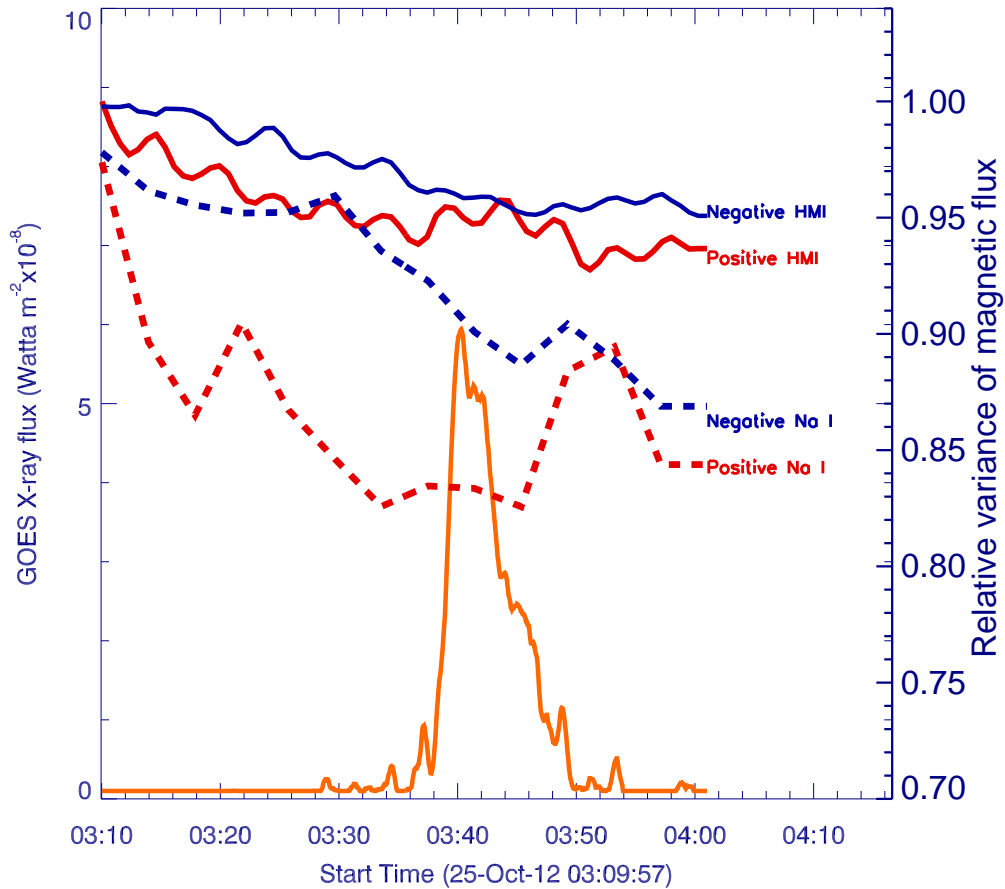}}
  % \caption{The magnetogram and 304\AA~image with sub-regions labeled 1, 2 and 3 to show the magnetic flux and radiation intensity in individual sub-regions.} \label{Fig4}
    \caption{The solid lines show the evolution of photospheric magnetic flux observed with 6301.2 Fe I line and the dotted line show the lower chromospheric magnetic flux observed with Na line, corresponds the subregion labeled by white circle in Fig 1. The flux of X-ray obtained from GOES is appended for the comparison of various phases of c-class flare.} \label{Fig3}
\end{figure}

\begin{figure}
%%%   G:\ynd\ytdata\20121025\hmi\readhmi-from-sav-hinode-pfnf-evolution
   \centerline{\includegraphics[width=1\textwidth,clip=]{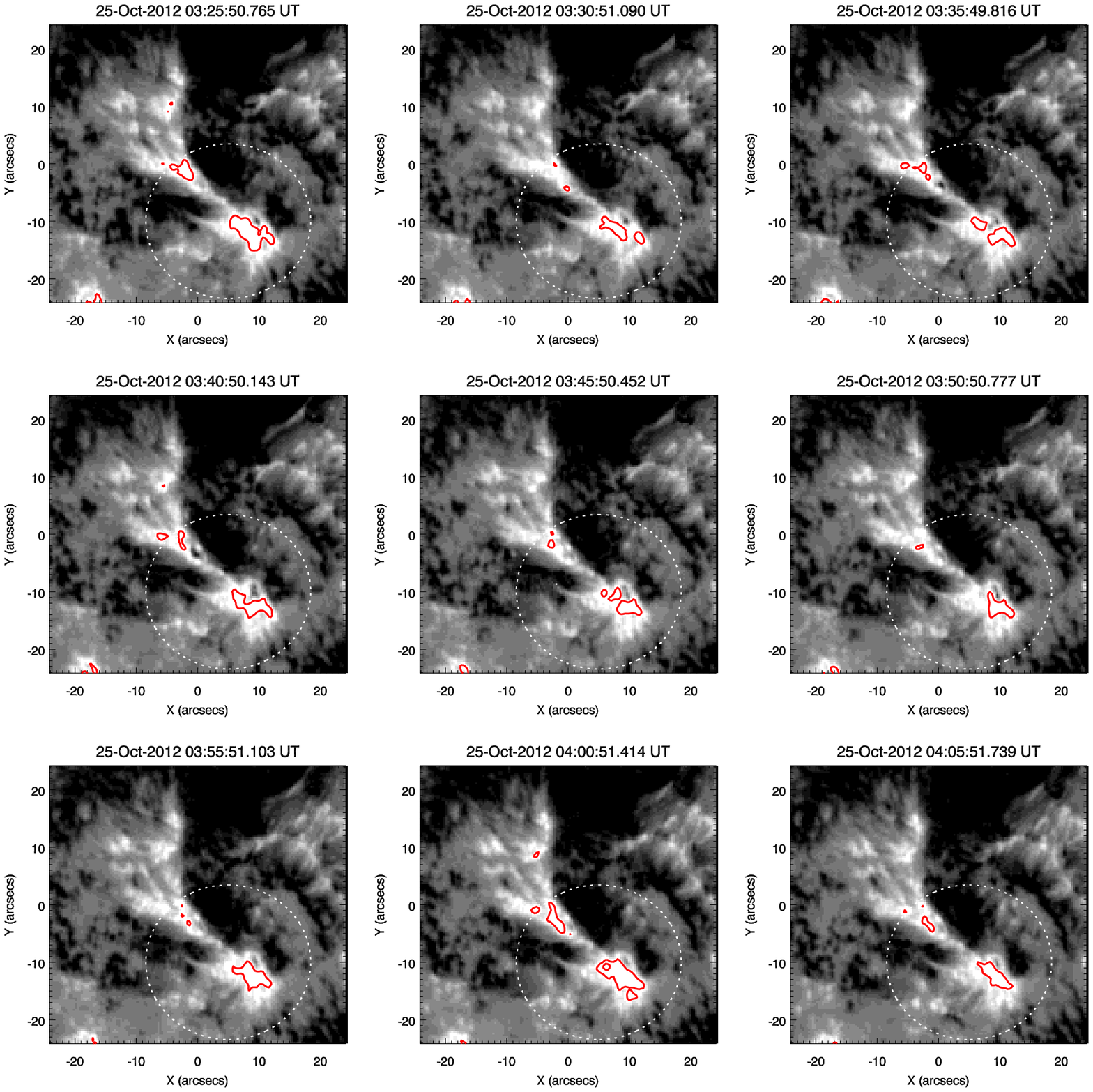}}
  % \caption{The magnetogram and 304\AA~image with sub-regions labeled 1, 2 and 3 to show the magnetic flux and radiation intensity in individual sub-regions.} \label{Fig4}
    \caption{The time series image of Na line with the corresponding contours obtained for SOT/$H$inode observations, the white circle drawn each panel is the region-of-interest where the inverse Evershed flows are identified same as figure 1.} \label{Fig4}
\end{figure}
%%%%%%%%%%%%%%%%%%%%%%%%%%%%%%%%%%%%%%%%%%%%%%%%%%%%%%%%%%%%%%%%%%%%%%%%%%%%%%%%%%%%%%%%%%%%%%%%%%%%%%%%%%%%%%%%%%%%%%%%%%%%%%%%%%%%%%%%%%%%%%%%%

In order to study the dynamical process of the event, we use the data obtained with the Atmospheric Imaging Assembly (AIA) and  Helioseismic and Magnetic Imager (HMI) on board the Solar Dynamics Observatory (SDO)~\citep{2012SoPh..275...17L, 2012SoPh..275..229S, 2012SoPh..275....3P}. The images in AIA, the 304 \AA{}, 171 \AA{} and 211 \AA{} channels with a resolution of 0.6$^{\prime\prime}$ pixel$^{-1}$ and the cadence of 12s are used for this study. The line-of-sight (LOS) magnetograms with a resolution of 0.6$^{\prime\prime}$ pixel$^{-1}$ and the cadence of 45s with HMI are used to study the magnetic field configuration of the flaring site and the surrounding area. The data processing is based on the standard Solar Softwares (SSW) related to these instruments (such as hmi$_{-}$prep.pro, aia$_{-}$prep.pro and drot.pro). For observations alignment, the image cross correlation is applied to match the features between the AIA 304 \AA{} and H$\alpha$, while AIA and HMI can be aligned through exact position and time observed information. Additionally, the high resolution G band 4304, Ca II and Na I observation obtained from SOT/H$inode$ is also used in this work \citep{kos07,
tsu08}. G band 4304 and Ca II H line data with high resolution of 0.21 arcsec/pixel can give the fine structures of photosphere and chromosphere, respectively. While Na I 5896 \AA line observation with resolution of 0.32 arcsec/pixel from SOT/H$inode$ can show the LOS magnetic field that slightly approach to the photosphere, which can used as a supplement for HMI magnetic field to display the structures and evolution of magnetic field interested. The H$inode$ data processing are also based on standard solar
software (SSW {\it e.g}, fg\_prep.pro), where dark subtraction, flat fielding, the correction of bad pixels and
cosmic-ray removal were done for filtergram images obtained by SOT.

%%%%%%%%%%%%%%%%%%%%%%%%%%%%%%%%%%%%%%%%%%%%%%%%%%%%%%%%%%%%%%%%%%%%%%%%%%%%%%%%%%%%%%%%%%%%%%%%%%%%%%%%%%%%%%%%%%%%%%%%%%%%%%%%%%%%%%%%%%%%%%%%%
\begin{figure}

%%%   jzhc_check_bzbt_together
   \centerline{\includegraphics[width=1\textwidth,clip=]{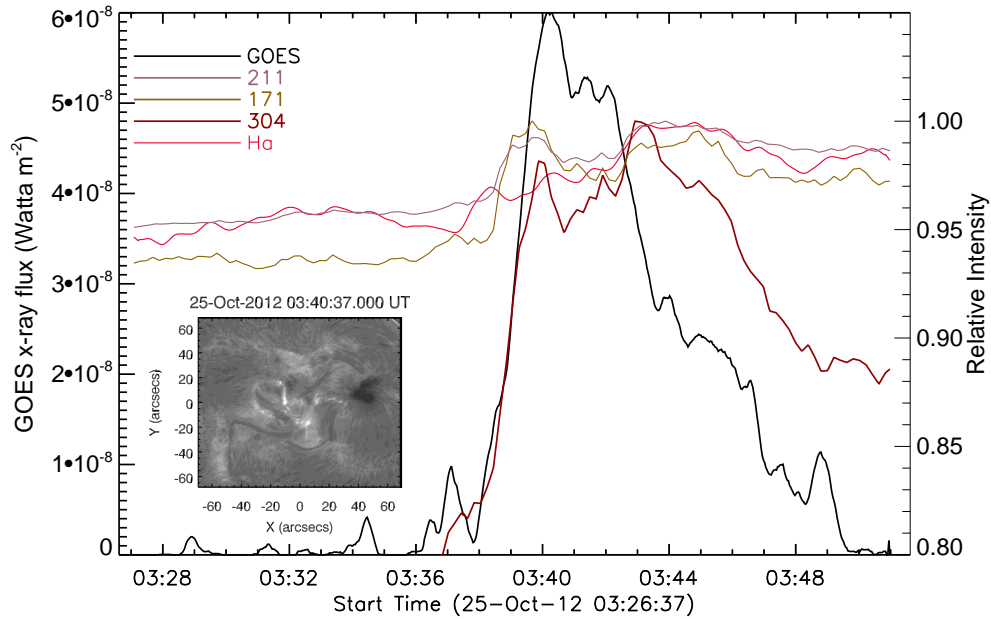}}
  % \caption{The magnetogram and 304\AA~image with sub-regions labeled 1, 2 and 3 to show the magnetic flux and radiation intensity in individual sub-regions.} \label{Fig4}
    \caption{It shows the evolution of  H$\alpha$ and GOES, AIA:304 \AA, 171 \AA, 211 \AA~ flux for the corresponding field of view labeled.} \label{Fig5}
\end{figure}
%%%%%%%%%%%%%%%%%%%%%%%%%%%%%%%%%%%%%%%%%%%%%%%%%%%%%%%%%%%%%%%%%%%%%%%%%%%%%%%%%%%%%%%%%%%%%%%%%%%%%%%%%%%%%%%%%%%%%%%%%%%%%%%%%%%%%%%%%%%%%%%%%

\begin{table}
\caption{Velocity of the propagation of brightness in  H$\alpha$ and AIA images
}
\label{T-simple}
\begin{tabular}{ccclc}     % define the column alignment
                           % l: left, c: center, r: right
  \hline                   % horizontal line
Slit Position & H$\alpha$ & AIA 304 nm & AIA 171 nm & AIA 211 nm \\
                    & km/s          & km/s           & km/s            & km/s           \\
  \hline
S1 & 69.52 & 62.50 & 61.44 & 61.83 \\
S2 & 99.50 & 90.62 &  93.98 & 84.58 \\
  \hline
\end{tabular}
\end{table}

\section{Observational Results}
\label{sect:Results}
Figure \ref{Fig1} shows the H$\alpha$ image and corresponding photospheric magnetogram of the sunspot and surrounding region. The encircled region in Figure 1a is the site of initial brightness of the flare, which is characterized by a narrow field structure of opposite polarity compared to their surrounding. The Figure 1 shows a quadrupolar magnetic configuration at the location of initial brightness of the flare that is shown by a circle on the SDO/HMI (middle panel (b)) and Na I (right panel (c)) snapshots. In the H$\alpha$ image, the bright fibrils of the superpenumbra are visible which seem to be terminated at the outer boundary of the positive polarity sunspot at its northern side. Figure \ref{Fig2} shows images of the same area in G-band and Ca II obtained by the Solar Optical Telescope onboard Hinode mission and the AIA 171 \AA image observed by SDO. The G-band and 171 \AA brightening in the location of flare and absence of Ca II structure suggest the presence of intense vertical magnetic field at this location.

Figure \ref{Fig3} shows the evolution of magnetic field near the flare location as a function of time.
Here the relative values of magnetic flux plotted which are normalized to initial time (03:10) in this plot.
At the mean time GOES X-ray flux is added to labeled flare process, and the positive and negative magnetic fluxes are calculated, respectively.
The flux cancellation prior to the flare onset and emergence around the flare time is clearly evident from the relative variance of positive and negative flux.
Figure \ref{Fig3} shows the evolution of the LOS magnetic field, where the flux seen to be decreasing unevenly indicating emergence and cancellation on the flare site. The circular polarization signals in Na line, that represent the magnetic flux in lower chromosphere also show a similar trend, except that there was an apparent imbalance of signal representing negative and positive field. The flare should be regarded as it was triggered by small scale magnetic structure highlighted by white circle plotted in Figure \ref{Fig1}, and the evident variances of this local magnetic field reflect its potentiality of trigger for this small fare and hereafter high speed plasma flow.
We might conclude the emerging flux to be tilted such that even at chromospheric height negative flux is seen enhanced. The flux cancellation continuously heated the localized plasma that moved along the super penumbral fibrils towards the sunspot. As the new flux emerge below the super penumbral canopy, it might reconnect with the pre-existing magnetic field lines above.
We describe such a situation in Figure \ref{Fig8} in a detailed manner in the forthcoming paragraphs. Additionally, to show the evolutions of magnetic fluxes more intuitively, the polarization signals in Na line with contours are shown in \ref{Fig4} basing on the time series observations, from the distributions of contours in \ref{Fig4} it can found that before flare there exist a evident decreases of magnetic fluxes.

%%%%%%%%%%%%%%%%%%%%%%%%%%%%%%%%%%%%%%%%%%%%%%%%%%%%%%%%%%%%%%%%%%%%%%%%%%%%%%%%%%%%%%%%%%%%%%%%%%%%%%%%%%%%%%%%%%%%%%%%%%%%%%%%%%%%%%%%%%%%%%%%%%
\begin{figure}
%%%   jzhc_check_bzbt_together
   \centerline{\includegraphics[width=1\textwidth,clip=]{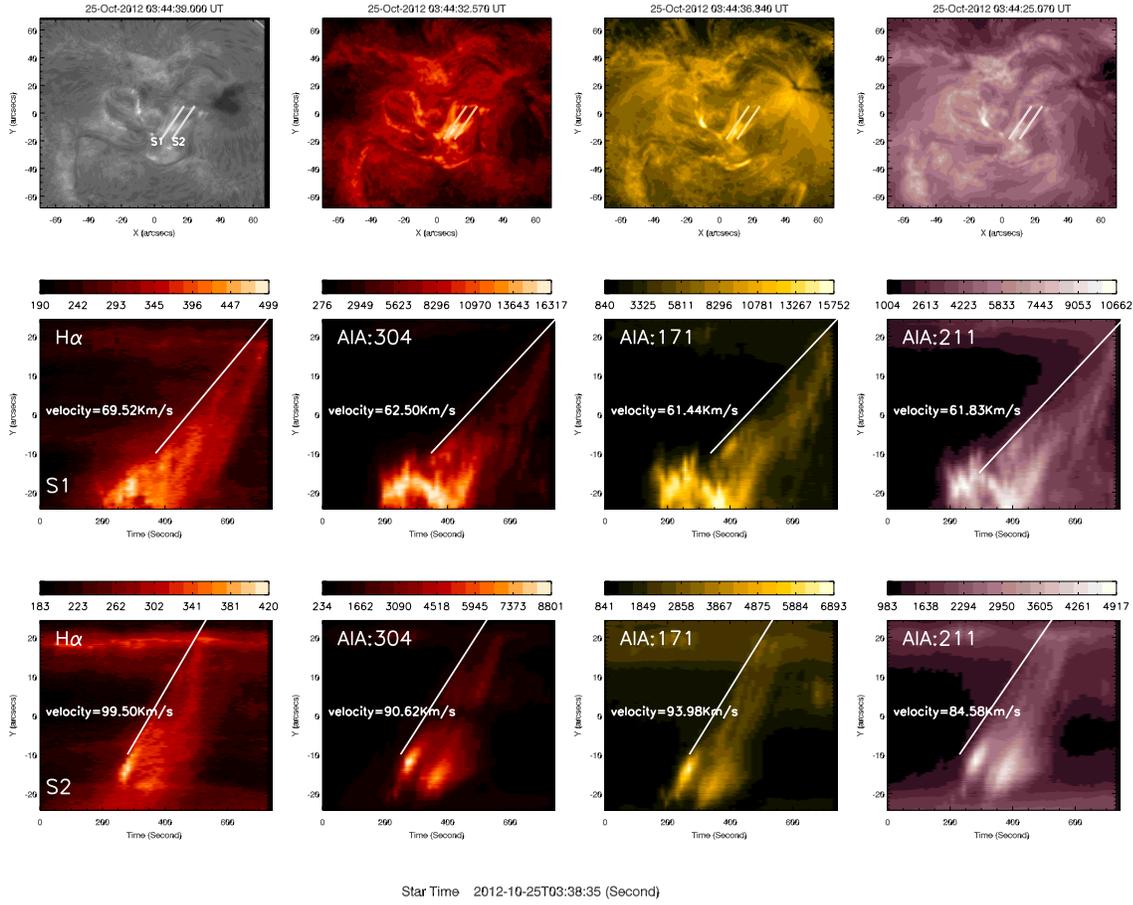}}
  % \caption{The magnetogram and 304\AA~image with sub-regions labeled 1, 2 and 3 to show the magnetic flux and radiation intensity in individual sub-regions.} \label{Fig4}
    \caption{H$\alpha$ and AIA:304 \AA, 171 \AA, 211 \AA~ time-distance plot for two slit selected from individual image, the slant
line used to show velocity labeled.} \label{Fig6}
\end{figure}

%%%%%%%%%%%%%%%%%%%%%%%%%%%%%%%%%%%%%%%%%%%%%%%%%%%%%%%%%%%%%%%%%%%%%%%%%%%%%%%%%%%%%%%%%%%%%%%%%%%%%%%%%%%%%%%%%%%%%%%%%%%%%%%%%%%%%%%%%%%%%%%%%%

%%%%%%%%%%%%%%%%%%%%%%%%%%%%%%%%%%%%%%%%%%%%%%%%%%%%%%%%%%%%%%%%%%%%%%%%%%%%%%%%%%%%%%%%%%%%%%%%%%%%%%%%%%%%%%%%%%%%%%%%%%%%%%%%%%%%%%%%%%%%%%%%%%
\begin{figure}
%%%   jzhc_check_bzbt_together
   \centerline{\includegraphics[width=1\textwidth,clip=]{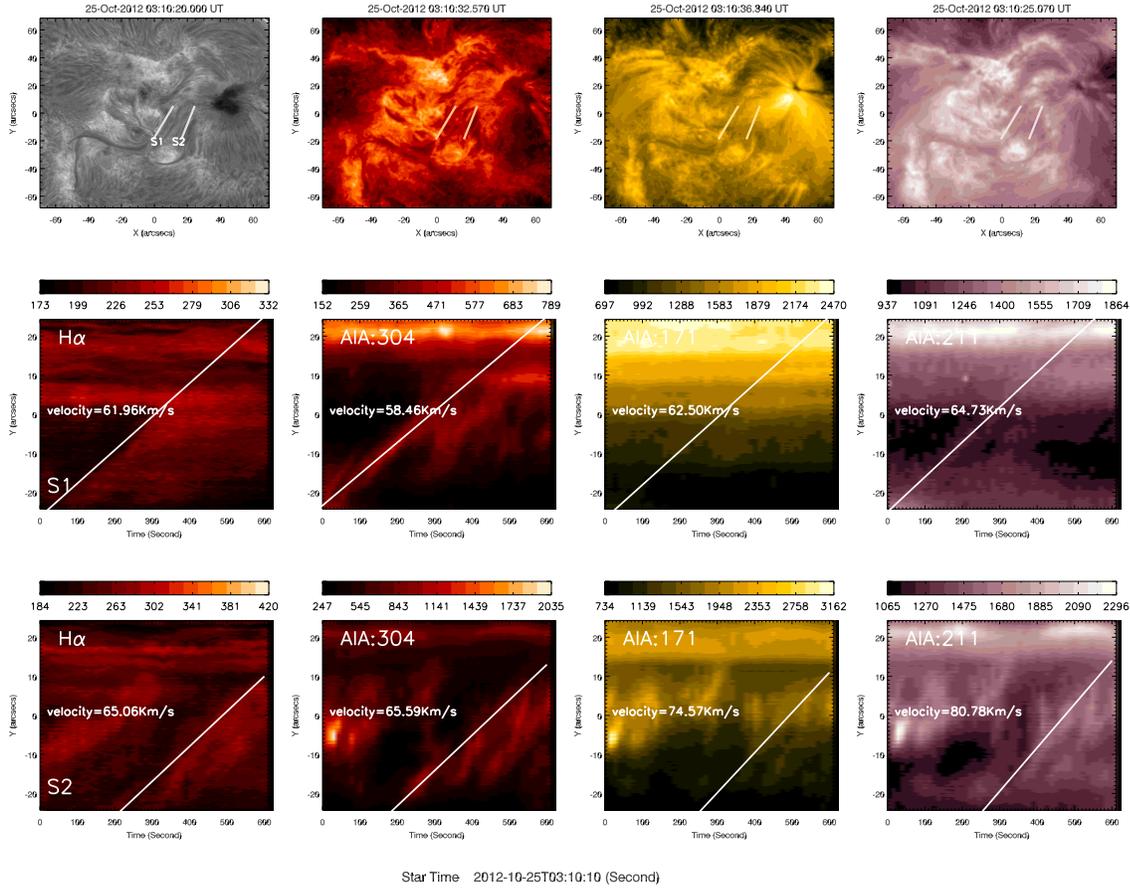}}
  % \caption{The magnetogram and 304\AA~image with sub-regions labeled 1, 2 and 3 to show the magnetic flux and radiation intensity in individual sub-regions.} \label{Fig4}
    \caption{H$\alpha$ and AIA:304 \AA, 171 \AA, 211 \AA~ time-distance plot for two slit selected from individual image obtained in quiet phase of the pre-flare stage, the slant line used to show velocity labeled.} \label{Fig7}
\end{figure}

%%%%%%%%%%%%%%%%%%%%%%%%%%%%%%%%%%%%%%%%%%%%%%%%%%%%%%%%%%%%%%%%%%%%%%%%%%%%%%%%%%%%%%%%%%%%%%%%%%%%%%%%%%%%%%%%%%%%%%%%%%%%%%%%%%%%%%%%%%%%%%%%%%

Figure \ref{Fig5} shows the light curve of the flaring region in GOES X-ray superimposed with H$\alpha$ and other wavelengths obtained by integrating the inserted field of view. The time difference between the peak of the light curve are less than 2 seconds. It is typical scenario of the compact flares (e.g., Benz, 2008; Shibata and Magara, 2011). In order to obtain the speed of the flare heated plasma, time-slice images are prepared by staking the images along two slits aligned to the super-penumbral fibrils as shown in Figure \ref{Fig6}. The speed of the brightness propagation in H$\alpha$ and AIA images obtained from the time-slice images are summarized in Table 1. The speeds in images in transition region images are about 10 km s$^{-1}$ lower than the speed in H$\alpha$ images. As  a reference in Figure \ref{Fig7}, we show pre-flare phase where such flows are absent.

Figure \ref{Fig8} describes the heating of chromospheric material and the propagation geometry in the form of a cartoon. The near simultaneous occurrence of light curve peaks in different wavelengths shown in Figure \ref{Fig3} suggest a low altitude reconnection event with the x-line shown in the figure, which is initially located in the chromosphere and then rapidly moves upward
(\citeauthor{2003Adv. Space. Res.?32..1043P}, \citeyear{2003Adv. Space. Res.?32..1043P}). %(Forbes, 2003).
Similar events resulting from low altitude have been observed with different magnitude have been observed with SMM
(\citeauthor{1990apj?73.137}, \citeyear{1990apj?73.137}).
%(Antonucci et al, 1990).

From the photospheric magnetic field configuration and appearance of initial brightening, it is clear that the plasma heating site was located near the outer edge of the super-penumbra. As x-point moved rapidly upward into chromosphere and transition region forming a current-sheet, the hot plasma got loaded on to the magnetic field line delineating super-penumbral fibrils and moved towards the penumbra in the same pattern as inverse-Evershed flow. If we consider the proper motion speeds of bright front as representative of mass motion of heated material, the speeds in H$\alpha$ are found to be around 60 km s$^{-1}$ to 90 km s$^{-1}$ and Transition region speeds are about 10 km s$^{-1}$ lower as can be seen in Table 1. These speeds are about 10 times higher than the normal Inverse-Evershed flow speeds.

%%%%%%%%%%%%%%%%%%%%%%%%%%%%%%%%%%%%%%%%%%%%%%%%%%%%%%%%%%%%%%%%%%%%%%%%%%%%%%%%%%%%%%%%%%%%%%%%%%%%%%%%%%%%%%%%%%%%%%%%%%%%%%%%%%%%%%%%%%%%%%%%%%
\begin{figure}
%%%   jzhc_check_bzbt_together
   \centerline{\includegraphics[width=1\textwidth,clip=]{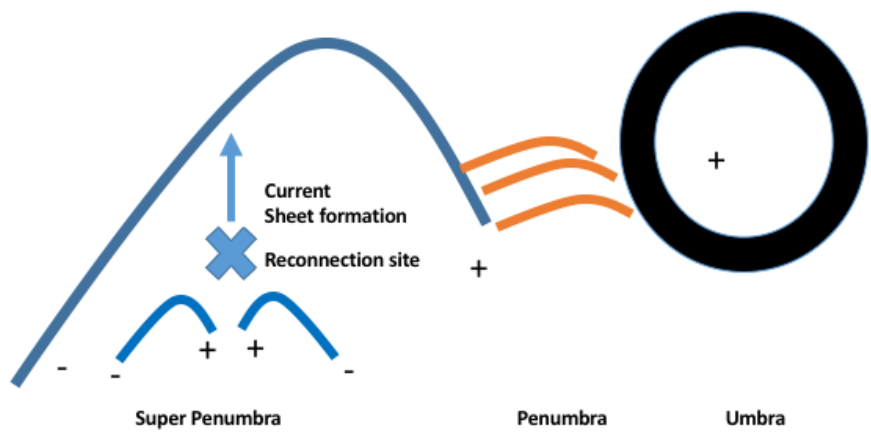}}
  % \caption{The magnetogram and 304\AA~image with sub-regions labeled 1, 2 and 3 to show the magnetic flux and radiation intensity in individual sub-regions.} \label{Fig4}
    \caption{Model of the event. The dark thick circle represent the positive polarity sunspot umbra. The maroon colored arches represent the penumbral fibrils.  Blue colored superpenumbral arch originate at the boundary of penumbral fibrils and extend to the mote region. The two blue arches below the superpenumbra represent the emerged flux that create an "X" point, which is a site of chromospheric reconnection. The hot plasma produced as a result of reconnection move upwards as shown by the arrow mark.} \label{Fig8}
\end{figure}

%%%%%%%%%%%%%%%%%%%%%%%%%%%%%%%%%%%%%%%%%%%%%%%%%%%%%%%%%%%%%%%%%%%%%%%%%%%%%%%%%%%%%%%%%%%%%%%%%%%%%%%%%%%%%%%%%%%%%%%%%%%%%%%%%%%%%%%%%%%%%%%%%%

When we analyze the inverse Evershed flows in the lower temperature band, the H$\alpha$ movie of the flare shows that the brightness moves from the heating site towards the sunspot and stops at the boundary marking the penumbra and super-penumbra. No brightness is observed in the penumbral region represented by the grey lines in the cartoon. It is remarkable to notice that even though the plasma moved at a speed about 10 times higher than the speed of inverse-Evershed material, they did not penetrate the umbral region as noticed in movies in H$\alpha$ and AIA wavelengths. This would imply that the super-penumbral fibrils carrying hot plasma terminate at the boundary where the umbral fibrils also dive into the sun and these two sets of magnetic field do not inter mingle. The barrier at this boundary is so strong that the high speed plasma can not over come.

\section{Discussion and Conclusions}

To the best of our knowledge, this is the first attempt to study the multi-tempretaure view of the inverse Evershed flows in and above the superpenumbral regions around a sunspot. As observed, during the onset of the flare, the speed of the heated plasma in H$\alpha$ is found to be around 60 km s$^{-1}$ to 90 km s$^{-1}$ and the speeds in Transition region and inner coronal layers are few tens of km s$^{-1}$ (Table 1). These speeds are about five times higher than the normal Inverse-Evershed flow speeds. In spite of large speeds even these flows are stopped at the boundary of the spot's penumbra likewise normal inverse Evershed flows.

Due to the different scale heights and density stratification in the solar atmosphere, these oppositely directed flows are having a less direct relationship with the photospheric Evershed effect (Teriaca et al., 2008). In fact, these dynamics are the set of the down flowing plasma from the top of the canopies in the superpenumbral region which may sometime be mixed with the chromospheric inverse Evershed flows.
At H$\alpha$ formation temperature the observed speed fall at supersonic scale between 60 - 90 km s$^{-1}$, which is larger than the typically observed speed of the inverse Evershed flows in the chromosphere (Schad et al., 2013). When, we study the flows detected at different temperatures, e.g., TR to inner corona in the same region, the flow speeds switch towards subsonic (Table 1). Therefore, multi temperature view of this flow is a mixture of the supersonic and subsonic plasma motions drifting towards the penumbra of the northward sunspot (Fig.~1 and 6). The higher speed of the inverse Evershed flows may be due to the modulation of flare energy release as well as changing magnetic field and plasma conditions at the site where they are originated.

Although the flows at multiple temperature plasma were ongoing, we also notice that the overlying loop-system visible at TR and inner coronal temperatures (Figure \ref{Fig6}) at the flare site on top of it (Figure \ref{Fig8}), exhibits the \textbf{possible} longitudinal oscillations with the period of approximately 300 s. The lower part of the slit position is lying (in projection) in such a manner obliquely that it detects the sloshing back and forth plasma on these tubes. Since, this behavior is not evident in the H$\alpha$ time-distance diagram, therefore, its origin is not in the lower solar atmosphere and it is most likely generated during the flare energy release in the overlying magnetic flux structure (Figure \ref{Fig8}). After repeating one cycle of the back and forth motion, finally the part of the plasma is started downflowing through the magnetic channels with almost same subsonic speed towards the penumbra of the sunspot lying northward. It should be noted that some plasma is already started flowing (the first linearly fitted envelope), while rest will flow with almost the same velocity after the oscillations (see the almost same slope in the in-bound envelope in Distance-time diagram).
However, the longitudinal oscillations maybe not the real oscillation signals, since only limited periods are detected. The alternative possibility is the longitudinal oscillations result from the repetitive flows triggered by continue magnetic reconnections.

We note an interesting point that in H$\alpha$, the plasma is instantly started flowing with the supersonic speed through the superpenumbral magnetic channel ($t$=200 s; first column in Fig.~5). However, in the TR/inner coronal bands, the part of the plasma flows in the same magnetic channel with a time lag of almost 200 s ($t$=400 s; second-forth column in Fig.~5). This indicates a very complex situation on the magnetic channel associated with the inverse Evershed effects, which may consist of the complex pattern of the flow of multi-temperature plasma with a range of downflow speeds. Moreover, they will also depend upon local magnetic field and plasma configuration and strength of the flare energy release (in any). Therefore, the multi-temperature and multi-height view of the inverse Evershed effect must be investigated in order to explore its underlying physical processes.

In spite of all complexities, the plasma flows towards the inverse Evershed channel over the penumbra of the sunspot. The discontinuity in the typical magnetic field and plasma properties at the adjoining of these two different sets of field lines (super-penumbra and spot's penumbra) further leads to the discontinuity in the characteristic magnetoacoustic and Alfv\'en speeds at the junction. This may further trap and stop the multi-temperature and mixed plasma flows what we observe in the present observational base-line. Future study will be devoted to identify the inverse Evershed flow channels associated with the variety of the flaring events, and constraining their multi-temperature and multi-height descriptions.

{\emph Acknowledgements:} One of the authors DPC gratefully acknowledges a travel grant from Chinese Academy
of Sciences to NAOC, where he completed a part of this work. The work funded by Chinese Academy of Science President's International Fellowship Initiative.
{\it Hinode} is a Japanese mission developed and launched by
ISAS/JAXA, collaborating with NAOJ as a domestic partner, NASA and
STFC (UK) as international partners. Scientific operation of the
{\it Hinode} mission is conducted by the {\it Hinode} science team
organized at ISAS/JAXA. The work was also partly supported by the Grants: 11203036, 11673033, 11427901, XDB09040200, 2014FY120300, 11373040, 11703042, the Key Laboratory of Solar Activity National Astronomical Observations, Chinese Academy of Sciences.

\label{lastpage}

\end{document}